\documentclass[twocolumn,showpacs,preprintnumbers,amsmath,amssymb]{revtex4}

\usepackage{graphicx}
\usepackage{dcolumn}
\usepackage{bm}

\begin{document}

\preprint{submitted to Phys. Rev. Lett.}

\title{
Magnetism of Cu${}_{6}$Ge${}_{6}$O${}_{18}$-xH${}_{2}$O ($x = 0 \sim 6$), a compound of the one-dimensional Heisenberg $S = 1/2$ model with competing antiferromagnetic interactions
}

\author{Masashi Hase}
 \email{HASE.Masashi@nims.go.jp}
\author{Kiyoshi Ozawa}
\author{Norio Shinya}

\affiliation{%
National Institute for Materials Science (NIMS), 1-2-1 Sengen, Tsukuba, 305-0047, Japan
}%

\date{\today}

\begin{abstract}

We measured the magnetic susceptibilities of Cu${}_{6}$Ge${}_{6}$O${}_{18}$-xH${}_{2}$O ($x = 0 \sim 6$). 
Susceptibility above the antiferromagnetic (AF) transition temperature ($T_{\rm N}$) agrees with susceptibility of the one-dimensional Heisenberg $S = 1/2$ model with competing AF interactions. 
From the estimated ratio between nearest-neighbor and next-nearest-neighbor AF exchange interactions, the spin system is probably located near a boundary between spin systems with gapless and gapped magnetic excitation. 
The value of $T_{\rm N}/T_{\rm max}$, where $T_{\rm max}$ is temperature at which the susceptibility is maximum, shows a unique dependence on $x$. 
The $x$ dependence can be explained qualitatively by taking phase transition as a function of $x$ and the effect of disorder into account.

\end{abstract}


\maketitle

Since the discovery of the spin-Peierls transition in CuGeO3 \cite{Hase93a}, this cuprate has attracted much attention and many works on this cuprate have been carried out \cite{Hase93b,Hase93c,Castilla95,Riera95,Sasago96,Masuda98}. 
Results of studies on CuGeO3 and related materials \cite{Sasago95} have indicated that some Ge or Si oxides including magnetic ions may have interesting spin systems. 

Among these oxides, we have been interested in Cu${}_{6}${\it M}${}_{6}$O${}_{18}$-xH${}_{2}$O ({\it M} = Ge or Si) because of their crystal structure. 
Cu${}_{6}$Si${}_{6}$O${}_{18}$-6H${}_{2}$O is a natural mineral called dioptase, and its crystal structure was determined in 1955 \cite{Heide55}. 
Cu${}_{6}$Ge${}_{6}$O${}_{18}$-$x$H${}_{2}$O was first synthesized in 1997 and was shown to have the same structure as that of Cu${}_{6}$Si${}_{6}$O${}_{18}$-$x$H${}_{2}$O \cite{Brandt97}. 
The value of $x$ can be changed in the range of 6 to 0 without change in the crystal structure by choosing conditions of thermal treatment \cite{Brandt97,Breuer88}. 
Atomic positions are shown schematically in Fig. 1 \cite{Ribbe77}. 
The space group is $R \bar 3$ (No. 148).
There are one Cu, one Ge, four O, and two H atom positions as a crystallographic view. 
Localized spins ($S = 1/2$) exist only on Cu${}^{2+}$ ions. 
In the case of $x = 6$, an octahedron surrounding a Cu atom is formed by two O(2), two O(3), and two O(4) atoms, where the O(4) atoms correspond to oxygen of water molecules. 
From Cu-O distances shown in TABLE I, we know that localized spins exist in $d_{x^{2}-y^{2}}$ orbits extending towards O(2) and O(3). 

In Cu${}_{6}$Ge${}_{6}$O${}_{18}$-6H${}_{2}$O, the distances between two Cu${}^{2+}$ ions are 2.96 and 3.27 \AA \ for the first-nearest-neighbor (1NN; pink bars) and second-nearest-neighbor (2NN; green bars) Cu-Cu bonds, respectively, while the distance is 4.93 \AA \ for the third-nearest-neighbor bond and is greater than the above two distances. 
Therefore, exchange interactions in the 1NN and 2NN Cu-Cu bonds ($J_1$ and $J_2$) should be taken into account. 
From the Cu-O-Cu angles shown in TABLE I, $J_2$ is positive (antiferromagnetic; AF) and the magnitude of $J_2$ is probably larger than that of $J_1$. 
It has been shown that $J_2$ is AF and $J_1$ is negative (ferromagnetic; F) by measurements of magnetic Bragg peaks below an AF transition temperature ($T_{\rm N}$) in Cu${}_{6}$Si${}_{6}$O${}_{18}$-xH${}_{2}$O with x = 0 or 6 \cite{Wintenberger93,Belokoneva02}. 
Each Cu has two 2NN bonds and one 1NN bond. 
Spiral chains are formed by the 2NN bonds and are connected to each other by the 1NN bonds. 
There is no magnetic competition in whatever sign of $J_1$ and $J_2$. 
Since the chains formed by the 2NN bonds are spiral, we may not be able to ignore exchange interaction ($J'_2$) in the next-nearest-neighbor (NNN) Cu-Cu bond in the chain (a green dotted bar) through the Cu-O(2)-O(2)-Cu path indicated by red bars in Fig. 1(b). 
If $J'_2$ is AF, magnetic competition occurs in the spiral chains.

Magnetic properties of Cu${}_{6}$Si${}_{6}$O${}_{18}$-xH${}_{2}$O with x = 0 or 6 have already been investigated. 
We can see a broad maximum around $T_{\rm max} \sim 45$ or 110 K in magnetic susceptibility of a powder sample with x = 6 or 0 \cite{Wintenberger93}. 
Gros et al. measured susceptibility of a single-crystalline sample with x = 6 and obtained $T_{\rm N}$ = 15.5 K \cite{Gros02}. 
The value of $T_{\rm max}$ suggesting magnitude of magnetic interactions in the compound with $x = 0$ is about two-times greater than that in the compound with $x = 6$ in spite of small quantitative changes in the crystal structures. 
Therefore, dependence of magnetic properties of Cu${}_{6}${\it M}${}_{6}$O${}_{18}$-$x$H${}_{2}$O on $x$ is interesting but has not been studied systematically. 
Thus, we investigated magnetic susceptibility of powder samples of Cu${}_{6}$Ge${}_{6}$O${}_{18}$-$x$H${}_{2}$O ($x = 0 \sim 6$). 

Crystalline powder of Cu${}_{6}$Ge${}_{6}$O${}_{18}$-6H${}_{2}$O was synthesized by reaction of copper (II) acetate  [Cu(CH$_3$COO)$_2$] and GeO${}_{2}$ in an aqueous solution \cite{Brandt97}. 
Samples with $x < 6$ were obtained by thermal treatment at 423 to 823 K in air. 
Values of $x$ were estimated by measuring the weights of samples before and after thermal treatment. 
Water is not removed when the temperature of the thermal treatment is lower than 373 K. 
Besides, once water has been removed, rehydration is impossible under ambient conditions. 
We obtained X-ray diffraction patterns at room temperature. 
As shown in Fig. 2, $a$ increases and $c$ decreases monotonically with increase in $x$. 
Structure parameters were refined by the Rietveld method from the X-ray diffraction data using the RIETAN program \cite{Izumi00} and are summarized in TABLE I. 
X-ray diffraction patterns of samples with $x = 0$ and 6 at 8.5 K were also obtained and were identical with those at room temperature. 
Magnetic susceptibility was measured using a superconducting quantum interference device magnetometer (Quantum Design MPMSXL). 
Besides, we confirmed that Cu${}_{6}$Ge${}_{6}$O${}_{18}$-$x$H${}_{2}$O is an insulator. 

\begin{table}
\caption{\label{table1}
Interatomic distances and angles in Cu${}_{6}$Ge${}_{6}$O${}_{18}$-$x$H${}_{2}$O. 
NNN means next-nearest neighbor {\it in the spiral chain}. 
}
\begin{ruledtabular}
\begin{tabular}{ccccc}
& $x = 0$ & 1.54 & 2.24 & 6\\
\hline
Cu-O(2) \AA & 1.95,1.96 & 1.91,1.97 & 1.93,1.95 & 1.97,1.99\\
Cu-O(3) \AA & 1.90,1.96 & 1.88,1.97 & 1.91,1.94 & 1.94,1.97\\
Cu-O(4) \AA &            & 2.41,2.68 & 2.53,2.57 & 2.58,2.65\\
Cu-Cu (1NN) \AA & 2.94 & 2.93 & 2.93 & 2.96\\
Cu-Cu (2NN) \AA & 3.26 & 3.26 & 3.27 & 3.27\\
Cu-Cu (NNN) \AA & 5.64 & 5.64 & 5.64 & 5.64\\
O(2)-O(2) (NNN) \AA & 2.86 & 2.85 & 2.84 & 2.86\\
Cu-O(3)-Cu (1NN) & $99.33^{\circ}$ & $99.18^{\circ}$ & $99.02^{\circ}$ & $98.45^{\circ}$\\
Cu-O(2)-Cu (2NN) & $112.45^{\circ}$ & $114.36^{\circ}$ & $114.53^{\circ}$ & $111.49^{\circ}$\\
\end{tabular}
\end{ruledtabular}
\end{table}

Figure 3(a) shows temperature ($T$) dependence of magnetic susceptibility $\chi (T)$ of Cu${}_{6}$Ge${}_{6}$O${}_{18}$-$x$H${}_{2}$O measured in magnetic fields of $H = 0.1$ T. 
In $\chi (T)$ of the sample with $x = 6$, we can see a broad maximum around $T_{\rm max} = 100$ K. 
The value of $T_{\rm max}$ increases with decrease in $x$. 
The broad maximum indicates that Cu${}_{6}${\it M}${}_{6}$O${}_{18}$-$x$H${}_{2}$O can be classified as low-dimensional antiferromagnets. 
This is consistent with the abovementioned notion that the spin system consists of spiral chains. 
A Curie-Weiss term is seen in $\chi (T)$ at low temperatures. 
We fitted $C/(T + \theta) + \chi_{0}$ to susceptibility below 6 K. 
Since the $T$ range used in this fitting was very narrow, susceptibility of Cu${}_{6}$Ge${}_{6}$O${}_{18}$-$x$H${}_{2}$O was assumed to be $T$-independent and was contained in the $T$-independent term, $\chi_{0}$. 
In the sample with $x = 6$, for example, we estimated $\theta$ to be 2.41 K and estimated the ratio of spins contributing to the Curie-Weiss term $p$ to be $2.00 \times 10^{-3}$ from the Curie constant $C$ assuming that the value of $g$ was 2.12. 
Similar values of $\theta$ and $p$ were obtained in other samples. 
In Fig. 3(a), $\frac{1}{1-p} [\chi (T) - C/(T + \theta)]$ of the sample with $x = 6$ is also shown by a dotted curve.
It should be noted that the Curie-Weiss term hardly affects determination of $T_{\rm max}$ and $T_{\rm N}$, and comparison between experimental and theoretical susceptibilities in Fig. 3(b).
The inset of Fig. 3(a) shows derivative curves [$d \chi (T)/ dT$]. 
In $d \chi (T)/ dT$ of the sample with $x = 6$ or 0, there is one peak at 38.5 or 73.5 K, which indicates phase transition to the AF long-range order (AFLRO). 
On the other hand, two transitions seem to overlap each other between 35 and 60 K in the sample with $x = 1.54$ \cite{Comment1}. 

We assumed that the orbital part of susceptibility was $1 \times 10^{-4}$ (emu/Cu mol) and subtracted that part from the experimental susceptibility. 
The remaining susceptibility $\chi_{\rm spin}$ times $T_{\rm max}$ is plotted as a function of $T/T_{\rm max}$ in Fig. 3(b). 
Taking experimental errors into account, we can say that these curves are similar to one another and show no systematic change as a function of $x$. 
Here, we compare these curves with theoretical ones obtained from the one-dimensional Heisenberg $S = 1/2$ model with competing AF interactions (competing model) whose Hamiltonian is ${\bf H} = \sum_{i} (J_2 {\bf S}_{i} \cdot {\bf S}_{i+1} + J'_2 {\bf S}_{i} \cdot {\bf S}_{i+2})$.
In this model, a spin gap opens between singlet ground and excited states when $\alpha (\equiv J'_2/J_2)$ exceeds a critical value of $\alpha_{c} = 0.2412$ \cite{Castilla95}. 
Therefore, a quantum phase transition occurs when $\alpha = 0.2412$. 
Theoretical curves with various values of $\alpha$ \cite{Kuroe03} and an experimental one of the sample with $x = 6$ are shown in the inset of Fig. 3(b). 
Since it is possible to compare experimental and theoretical curves in a wide $T/T_{\rm max}$ range, we use the curve of $x = 6$. 
The experimental curve agrees with the theoretical curves with $\alpha = 0.25$ and $g = 2.05$ or with $\alpha = 0.29$ and $g = 2.12$, while it does not agree with the theoretical curves with $\alpha = 0.20$ and 0.34. 
This agreement indicates that the spin system in Cu${}_{6}$Ge${}_{6}$O${}_{18}$-$x$H${}_{2}$O is located near a boundary between spin systems with gapless and gapped magnetic excitation, although $\alpha$ may depend weakly on $x$. 
The same conclusion was obtained even when the value of the orbital part of susceptibility was changed from 1 to $2 \times 10^{-4}$ (emu/Cu mol). 
Of course, there is a clear difference between the competing model and Cu${}_{6}$Ge${}_{6}$O${}_{18}$-$x$H${}_{2}$O. 
The ground state is a spin singlet in the model, while it is AFLRO in Cu${}_{6}$Ge${}_{6}$O${}_{18}$-$x$H${}_{2}$O due to interchain interaction ($J_1$). 
It is known that there are a few compounds that have both NN and NNN interactions in chains, such as CuGeO${}_{3}$ \cite{Castilla95,Riera95}, CaV${}_{2}$O${}_{4}$ ($S = 1$) \cite{Kikuchi01}, and NaV(WO${}_{4}$)${}_{2}$ ($S = 1$) \cite{Masuda02}. 

The dependence of $T_{\rm max}$ on $x$ is shown in Fig. 4(a), and the dependence of $T_{\rm N}/T_{\rm max}$ on $x$ is shown in Fig. 4(b). 
As $x$ decreases, $T_{\rm max}$ and therefore magnitude of exchange interactions increase monotonically. 
The value of $T_{\rm N}/T_{\rm max}$ decreases with decrease in $x$ from $x = 6$, while it is nearly constant below $x = 1.54$. 
At $x = 1.54$, $T_{\rm N}/T_{\rm max}$ jumps abruptly from 0.23 to 0.39, and two AF transitions are seen. 
The jump indicates that AFLRO at $x > 1.54$ and $x < 1.54$ belongs to essentially different phases. 
The two AF transitions at $x = 1.54$ are due to a phase separation into high and low $x$ phases. 
A phase separation always appears in the case of a first-order phase transition.
The jump is reminiscent of Mg concentration dependence of $T_{\rm N}$ in Cu${}_{1-y}$Mg${}_{y}$GeO${}_{3}$ \cite{Masuda98}. 
The value of $T_{\rm N}$ changes abruptly at a critical concentration $y_{c} = 0.023$. 
Lattice dimerization exists and affects AFLRO at $y < y_{c}$, while AFLRO appears in a uniform lattice at $y > y_{c}$.

Let us now discuss the dependence of $T_{\rm max}$ on $x$. 
The change in $T_{\rm max}$ cannot be attributed to only Cu-Cu distances and Cu-O-Cu angles because these values are almost independent of $x$ (TABLE I). 
In a Cu-octahedron, $\pi$-bonds are formed between Cu and O(4) of a water molecule indicated by black dotted bars in Fig. 1(b) \cite{Belokoneva02}. 
Thus, electron distribution in the Cu-octahedron is probably changed by extraction of water. 
Therefore, it is considered that exchange of spins in the Cu-O(2) and Cu-O(3) bonds increases with decrease in $x$, resulting in increase in exchange interactions, and therefore $T_{\rm max}$. 

Next, let us discuss the dependence of $T_{\rm N}/T_{\rm max}$ on $x$. 
Based on the results indicating that the spin system in Cu${}_{6}$Ge${}_{6}$O${}_{18}$-$x$H${}_{2}$O is probably located near a boundary between spin systems with gapless and gapped magnetic excitation, we assume that a spin gap due to competing AF interactions exists at $x < 1.54$ and that magnetic excitation is gapless at $x > 1.54$. 
Of course, the existence of a spin gap has not been confirmed in the susceptibility. 
However, it should be emphasized that powder-averaged susceptibility of CuWO${}_{4}$ does not clearly show existence of a spin gap due to appearance of AFLRO at 24 K \cite{Anders72}, although this cuprate has a gap of 1.4 meV (16 K) at the magnetic zone center  originating in alternation of AF exchange interactions in $S=1/2$ chains \cite{Lake96}.
It should also be emphasized that a spin gap and AFLRO can coexist. 
Examples are doped CuGeO${}_{3}$ \cite{Sasago96}, CuWO${}_{4}$ \cite{Anders72,Lake96}, CsNiCl${}_{3}$ ($S = 1$) \cite{Buyers86}, and SrNi${}_{2}$V${}_{2}$O${}_{8}$ ($S = 1$) \cite{Zheludev00}. 

Disorder is introduced by inhomogeneity of water in samples with $0 < x < 6$ and has a destructive effect on AFLRO. 
Thus, $T_{\rm N}/T_{\rm max}$ decreases with decrease in $x$ at $x >1.54$. 
Disorder also has a destructive effect on a spin gap. 
The destruction of a spin gap leads to development of AF correlation and therefore leads to increase in $T_{\rm N}/T_{\rm max}$.
Thus, as $x$ increases at $x < 1.54$, decrease in $T_{\rm N}/T_{\rm max}$ due to the destructive effect on AFLRO compensates increase in $T_{\rm N}/T_{\rm max}$ due to the destructive effect on a spin gap. 
As a result, $T_{\rm N}/T_{\rm max}$ is almost independent of $x$ at $x < 1.54$. 
Note that $T_{\rm N}$ increases with increase in dopant concentration at a low concentration in spin systems that have a singlet ground state and a spin gap between singlet and excited states, such as CuGeO${}_{3}$ \cite{Hase93b,Masuda98}, $S = 1/2$ two-leg-ladder SrCu${}_{2}$O${}_{3}$ \cite{Azuma97}, and $S = 1$ Haldane-material PbNi${}_{2}$V${}_{2}$O${}_{8}$ \cite{Uchiyama99}. 
In these cases, it is considered that development of AF correlation due to destruction of a spin gap is more dominant than the destructive effect of disorder on AFLRO. 
In order to prove that the abovementioned idea in Cu${}_{6}$Ge${}_{6}$O${}_{18}$-$x$H${}_{2}$O is correct, we have to confirm the existence of a spin gap. 
Therefore, it is necessary to make single crystals of Cu${}_{6}$Ge${}_{6}$O${}_{18}$-$x$D${}_{2}$O and to measure neutron scattering. 
Since $T_{\rm N}$ is not so small in comparison with $T_{\rm max}$, $J_{1}$ is not negligible. 
Thus, in order to determine $\alpha$ more accurately, we will use theoretical susceptibility obtained from the competing model taking $J_{1}$ into account, which has not been reported so far. 

We are grateful 
to K. Uchinokura, A. Tanaka, and T. Hikihara 
for their valuable discussions, 
to A. P. Tsai, T. Sato, H. Mamiya, and H. Suzuki 
for susceptibility measurements, 
to J. Wang and J. Ye 
for Rietveld analysis, 
to T. Furubayashi and I. Nakatani 
for X-ray diffraction measurements at low temperature, and 
to H. Kuroe 
for data of susceptibility obtained from the competing model. 
This work was supported by a grant for Intelligent Material Research from the Ministry of Education, Culture, Sports, Science, and Technology of Japan.

\newpage 

\begin{figure} 
\caption{
(color)
(a) 
Schematic drawing of the crystal structure of Cu${}_{6}$Ge${}_{6}$O${}_{18}$-$x$H${}_{2}$O.
Pink and green bars indicate the 1NN and 2NN Cu-Cu bonds. 
(b) 
Nearest-neighbor Cu-Cu and Cu-O bonds. 
Solid pink and green bars indicate the 1NN and 2NN Cu-Cu bonds, and the dotted green bar indicates the NNN Cu-Cu bond in the spiral chain. 
The red dotted bar represents an O(2)-O(2) bond in a Cu-O(2)-O(2)-Cu path responsible for the formation of the NNN Cu-Cu bond in the spiral chain. 
In both figures, sizes of spheres are smaller than ionic radii in this scale in order to avoid complexity. 
}
\label{Fig1}
\end{figure}

\begin{figure} 
\caption{
Dependence of lattice parameters of Cu${}_{6}$Ge${}_{6}$O${}_{18}$-$x$H${}_{2}$O on $x$. 
Closed and open circles indicate the data obtained in this work and from Ref. \cite{Brandt97}, respectively. 
}
\label{Fig2}
\end{figure}

\begin{figure} 
\caption{
(a) 
Temperature dependence of magnetic susceptibility $\chi (T)$ of Cu${}_{6}$Ge${}_{6}$O${}_{18}$-$x$H${}_{2}$O. 
The values of $x$ are 6, 5.43, 3.95, 2.90, 2.24, 1.54, 1.08, 0.80, and 0 from the upper to lower solid curves. 
The dotted curve represents $\frac{1}{1-p} [\chi (T) - C/(T + \theta)]$ of the sample with $x = 6$.
The inset shows $d \chi (T)/dT$ curves of the samples with $x = 6$, 1.54, and 0. 
One division of the vertical scale means $5 \times 10^{-6}$ (emu/Cu mol K). 
The curve of $x = 0$ has been shifted vertically by $2 \times 10^{-6}$ (emu/Cu mol K).
The two arrows indicate AF transition temperatures in the sample with $x = 1.54$. 
(b) 
$\chi_{\rm spin} T_{\rm max}$ versus $T/T_{\rm max}$, where $\chi_{\rm spin}$ is defined as $\chi (T) - 1 \times 10^{-4}$ (emu/Cu mol). 
Each curve in (b) corresponds to the curve with the same symbols in (a). 
The inset shows the experimental $\chi_{\rm spin} T_{\rm max}$ curve in the sample with $x = 6$ ($\bullet$) and theoretical ones of the competing model. 
The parameters are $\alpha = 0.20$ and $g = 2$ ($\bigcirc$), $\alpha = 0.25$ and $g = 2.05$ ($\square$), $\alpha = 0.29$ and $g = 2.12$ ($\diamondsuit$), and $\alpha = 0.34$ and $g = 2.25$ ($\triangle$). 
}
\label{Fig3}
\end{figure}

\begin{figure} 
\caption{
Dependence of $T_{\rm max}$ and $T_{\rm N}/T_{\rm max}$ of Cu${}_{6}$Ge${}_{6}$O${}_{18}$-$x$H${}_{2}$O on $x$.
}
\label{Fig4}
\end{figure}

\end{document}